\documentstyle[amssymb,preprint,aps]{revtex}

\newtheorem{theorem}{Theorem}
\newtheorem{acknowledgement}[theorem]{Acknowledgement}

\begin{document}
\title{Symmetry Analysis of ZnSe(100)/Air Interface By Second Harmonic Generation}
\author{Xiangyang Song$^{1}$, Arnold Neumann, Rein Maripuu, Wolfgang Seibt, and Kai
Siegbahn$^{\ast }$}
\address{ESCA-Laser Lab., Uppsala University, S-751 21, Uppsala, Sweden\\
$^{1}$Shanghai Institute of Optics and Fine Mechanics, \\
P. O. Box 800-211, P. R.China}
\maketitle

\begin{abstract}
We measured the polarized and azimuthal dependencies of optical second
harmonics generation (SHG) on polished surfaces of ZnSe(100) single crystal
surface in air, using a fundamental wavelength of 1.06$\mu m$. By
considering both, the bulk- and surface- optical nonlinearities within the
electric dipole approximation, we analysed the data for all four combination
of $p$- and $s$-polarized incidence and output. The measurement using $%
S_{in}-S_{out}$ is thereby particularly useful in the determining of the
symmetry ZnSe(100)/contamination layer-interface, which would lower the
effective symmetry of ZnSe(100) from $C_{4v}$ to $C_{2v}.$ The analysis of $%
p $-incident and $p$-output configuration allows us to distinguish the
[011]- and [0$\overline{1}$1]- directions.
\end{abstract}

\section{INTRODUCTION}

Optical Second Harmonic Generation(SHG) has matured to be an effective
surface- and interface-sensitive technique for probing a large variety of
surface- and interface-properties. This technique is a non-destructive. It
can be used for {\it in-situ} measurements in almost any environment,
provided that we have optical access to the sample \cite{Lupcke99, Shen97,
Reider95}. The method bases on the idea that generally a surface and a bulk
have different structural symmetry\cite{Shen97, Shen84, Stehlin88, Shen99}.
Materials with inversion symmetry, such as Si- and Ge- single crystals have
no bulk electric-dipole contribution. Electric quadrupole and magnetic
dipole contributions originate the leading source in SHG from the bulk.
However, a surface or an interface breaks the inversion symmetry and
produces an electric-dipole contribution. Generally the SHG-intensity from a
surface or an interface can be compared to be much greater than that from
the bulk, depending on the material, the photon energy, and the geometry of
the experiment. On the other hand, noncentrosymmetric materials like
crystals with the zincblende structure as, e.g. the compound semiconductors
GaAs and ZnSe, have intense bulk electric-dipole contributions to the SHG,
thus hampering surface observations. Stehlin et al \cite{Stehlin88}
illustrated possible combinations of the polarization of incident and SHG
output light and of crystal azimuths for which only the surface contribution
may be deduced for typical low index surface, e.g. (001), (110), and (111)
and used SHG as effective surface probe with submonolayer sensitivity to
monitor the adsorption of Sn on GaAs(001) surface. Later, S. R. Armstrong et
al \cite{Armstrong92} investigated by means of SHG a GaAs(100)/air
interface. Similar experiments performed by C. Yamada and T. Kimura \cite
{Yamada93, Yamada94} firstly observed the twofold rotational angledependence
(rotation anisotropy) of SHG in the reflected light from a well prepared and
by RHEED characterised noncentrosymmetric single crystal surface of
GaAs(100). They interpreted it as the interference of surface SHG\ and
dipole-allowed bulk SHG having fourfold symmetry results in twofold
anisotropy, whereby the degree of rotational anisotropy monitors the surface
reconstruction. Galeckas et al \cite{Galeckas92} showed that SHG in
reflection is a sensitive and practical method to invesigate the departure
from perfect crystalline ofder at the surface by studying the SHG rotational
anisotropy of crystalline, polycrystalline and amorphous silicon carbide
surfaces. Bottomley et al \cite{Bottomley93} determined to within $\pm 0.1%
{{}^\circ}%
$ the orientation of vicinal GaAs(001) and Si(111) single crystal wafers
using second and third harmonic generation. M. Takebayash et al \cite
{Takebayashi97} demonstrated the SHG-measurement of the tilt angle $\theta $
of the crystallographic axis of a vicinal GaAs(100) wafer towards a
direction $\xi $ that characterises step direction and step height and forms
an angle $\psi $ with respect to [100]-axis and found that the $s$-incident
polarization is useful in the determination of this tilt angle. C. Jordan et
al \cite{Jordan97} used the azimuthal rotational anisotropy of the SHG for 
{\it in-situ} finger print characterisation of various polytypes of SiC and
compared with X-ray diffraction (XRD) and identify defect regions in big
samples by observating the spatially resolved dependence of the SH-intensity.

In recent years, there has been big interest in wide-band gap II-VI compound
semiconductors. In particular, ZnSe is being actively studied due to their
electroluminescent properties in blue-light-emitting laser diodes \cite
{Haase}. These devices are fabricated by epitaxial growth on GaAs
substrates. The replacement of GaAs substrates by ZnSe ones may decrease the
concerntration of defects and increase the lifetime of the device. This can
be one possible way to improve ZnSe-based laser characteristics. An recent
observation in ZnSe/ZnCdSe laser diodes grown onto ZnSe substrates indicates
that the growth rate of the defects is much smaller than for layers grown
onto GaAs substrates \cite{Ohno98}. For the successful growth of II-VI
epilayers onto ZnSe layers is very importment to have well ordered substrate
surfaces. However, the superiority of ZnSe substrates over GaAs one has not
been realised yet, mainly because it is difficult to obtain high-quality
ZnSe substrates.

SHG has been successfully applied to the study of semiconductor surfaces
semiconductor/semiconductor heterostructures and interfaces, and
semiconductor-oxide interfaces \cite{Lupcke99}

and provides a powerful tool for understanding the energetics of various
defect formations and surface reconstructions. Although these experiments
have been performed mostly for GaAs surface \cite
{Stehlin88,Armstrong92,Yamada93,Yamada94} or SiC surface \cite
{Galeckas92,Jordan97}, similar symmetric studies were not extend to many
other non-centrosymmetric materials, such as ZnSe.

In this paper, we have measured the SH intensity from ZnSe(100) as a
function of the light polarization and as a function of the azimuthal angle.
We have also calculated the intensity of SHG from the surface electric
dipoles and the the second-order nonlinear polarization arising from the
bulk response. We use optical SHG to determine the macroscopic symmetry
properties of the ZnSe(100)/air interface. The remainder of the paper is
organized as follows. In Sec. II, the theory for SHG in reflection from the
bulk and surface of nonsymmetric crystals is briefly introduced. Then the
experimental apparatus is described in Sec. III, and finally the results are
analyzed in terms of symmetry.

\section{THEORY}

\subsection{General}

We can express the induced nonlinear surface polarization at an interface in
the following manner: \cite{Stehlin88} 
\begin{equation}
{\bf P}_{eff}(2\omega )={\bf P}_{surf}+{\bf P}_{bulk}F(\omega )F(2\omega
)L_{eff},
\end{equation}
where $F(\omega )$ and $F(2\omega )$ are the Fresnel factors for the
incident input and output fields, and $L_{eff}=(W+2w)^{-1}$ is the effective
phase-matching distance of the substrate with $W=k_{\omega ,z}$ and $%
2w=k_{2\omega ,z}$ representing the $z$ component of the wave vectors of the
fundamental and SH light, resepectively, ${\bf P}_{surf}$ is the surface
nonlinear polarization of the interfacial layer, and ${\bf P}_{bulk}$ is the
bulk nonlinear polarization in medium.

The bulk nonlinear polarization can generally be expressed by a series of
multipole terms: \cite{Guyot88} : 
\begin{equation}
{\bf P}_{bulk}(2\omega )={\bf P}^{(2)}(2\omega )-\nabla \cdot {\bf Q}%
^{(2)}(2\omega )+\frac{c}{2i\omega }\nabla \times {\bf M}^{(2)}(2\omega ),
\label{Eq-a}
\end{equation}
where ${\bf P}^{(2)},{\bf Q}^{(2)},$and ${\bf M}^{(2)}$ describe the
electric-dipole polarization, electric-quadrupole polarization and
magnetic-dipole polarization respectively. Up to the first derivative in $%
{\bf P}_{eff}^{(2)}$, the three polarization sources are the following: 
\begin{equation}
{\bf P}^{(2)}(2\omega )={\bf \chi }^{D}:E(\omega )E(\omega )+{\bf \chi }%
^{P}:E(\omega )\nabla E(\omega ),  \label{Eq-b}
\end{equation}
\begin{equation}
{\bf Q}^{(2)}(2\omega )={\bf \chi }^{Q}:E(\omega )E(\omega ),
\end{equation}
\begin{equation}
{\bf M}^{(2)}(2\omega )={\bf \chi }^{M}:E(\omega )E(\omega ).
\end{equation}
We neglect ${\bf M}$ and higher-order multipole in our discussion here,
because these contributions are thought to be orders of magnitude smaller
than the electric dipolar contribution. The second terms in eq. (\ref{Eq-a})
and (\ref{Eq-b}) arises from the gradient of the field $E(\omega )$. Since
the penetration depth in crystals ($\thicksim $100\AA ) is much smaller than
the spatial variation of the field ($\thicksim $10,000\AA\ for optical
frequencies), this bulk contribution is small under nonresonant conditions. 
\cite{Richmond95}

The surface contribution of the SH field arises from two effects. First, at
the interface between the dissimilar media, inversion symmetry is broken for
centrosymmetric crystals and symmetry is changed for noncentrosymmetric
crystals, and so a dipolar contribution to SHG can exist. In addition, there
is a discontinuity in the fundamental electric field normal to the surface.
This can generate a sizable SH contribution through higher-order multipole
terms. So the surface contribution to the nonlinear polarization can be
expressed as: 
\begin{equation}
P_{surf}^{(2)}(2\omega )=\chi ^{D}:E(\omega )E(\omega )+\chi ^{D}:E(\omega
)\nabla E(\omega )-\chi ^{Q}:\nabla E(\omega )E(\omega ).  \label{Eq-c}
\end{equation}
Since the tangential components of the incident electric field are
continuous across the interface, all terms in eq. (\ref{Eq-c}) involving a
gradient that can be neglected when describing an in-plane surface response
and only the first term should be considered: 
\begin{equation}
P_{surf}^{(2)}(2\omega )=\chi ^{D}:E(\omega )E(\omega ).
\end{equation}

If the nonlinear susceptibilities are all defined in terms of the input
field ${\bf E}(\omega )$, in the interfacial layer by ${\bf P}(\omega )={\bf 
{\chi }}^{(2)}:E(\omega )E(\omega )$, then from (\ref{Eq-a}), the
corresponding effective surface nonlinear susceptibility is given by 
\begin{equation}
{\bf \chi }_{eff}^{(2)}={\bf \chi }_{surf}^{(2)}+{\bf \chi }%
_{bulk}^{(2)}L_{eff}F(\omega )F(2\omega ),
\end{equation}
where ${\bf \chi }_{surf}^{(2)}$ and ${\bf \chi }_{bulk}^{(2)}$ are the
surface and bulk nonlinear susceptibility tensors, respectively. The
tensorial properties of ${\bf \chi }_{eff}^{(2)}$ can be exploited as: if
the SH intensity is recorded as a function of azimuthal angle of rotation,
the variation in intensity reflects the overall symmetry of the surface of
interface.

A tensorial expression for the second-order polarization can be written as: 
\cite{Richmond95}

\begin{equation}
\left( 
\begin{array}{c}
P_{x}(2\omega ) \\ 
P_{y}(2\omega ) \\ 
P_{z}(2\omega )
\end{array}
\right) =\left( 
\begin{array}{cccccc}
\chi _{xxx} & \chi _{xyy} & \chi _{xzz} & \chi _{xyz} & \chi _{xxz} & \chi
_{xxy} \\ 
\chi _{yxx} & \chi _{yyy} & \chi _{yzz} & \chi _{yyz} & \chi _{yxz} & \chi
_{yxy} \\ 
\chi _{zxx} & \chi _{zyy} & \chi _{zzz} & \chi _{zyz} & \chi _{zxz} & \chi
_{zxy}
\end{array}
\right) \times \left( 
\begin{array}{c}
E_{x}(\omega )E_{y}(\omega ) \\ 
E_{y}(\omega )E_{y}(\omega ) \\ 
E_{z}(\omega )E_{z}(\omega ) \\ 
2E_{y}(\omega )E_{z}(\omega ) \\ 
2E_{x}(\omega )E_{z}(\omega ) \\ 
2E_{x}(\omega )E_{y}(\omega )
\end{array}
\right) ,  \label{Eq-d}
\end{equation}
This is the general expression describing the interaction of the two EM
driving fields being coupled throuth a dyadic product.

\subsection{Bulk contribution}

We now define a new set of coordinates $(\widehat{{\bf x}}^{\prime },%
\widehat{{\bf y}}^{\prime },\widehat{{\bf z}}^{\prime })$ for each of the
three crystal orientations such that $z$ axis is perpendicular to each
crystal face. For the (111) crystal face, we therefore have, in terms of the
standard crystal axes $(\widehat{{\bf x}},\widehat{{\bf y}},\widehat{{\bf z}}%
)$ 
\begin{equation}
\left( 
\begin{array}{c}
\widehat{{\bf x}}^{\prime } \\ 
\widehat{{\bf y}}^{\prime } \\ 
\widehat{{\bf z}}^{\prime }
\end{array}
\right) =R_{1}\left( 
\begin{array}{c}
\widehat{{\bf x}} \\ 
\widehat{{\bf y}} \\ 
\widehat{{\bf z}}
\end{array}
\right) =\left( 
\begin{array}{ccc}
2/\sqrt{6} & -1/\sqrt{6} & -1/\sqrt{6} \\ 
0 & 1/\sqrt{2} & -1/\sqrt{2} \\ 
1/\sqrt{3} & 1/\sqrt{3} & 1/\sqrt{3}
\end{array}
\right) \left( 
\begin{array}{c}
\widehat{{\bf x}} \\ 
\widehat{{\bf y}} \\ 
\widehat{{\bf z}}
\end{array}
\right) ,
\end{equation}
where the new $\widehat{{\bf x}}^{\prime }$ axis is projected on to the
original crystal $\widehat{{\bf x}}$ axis in the plane of the crystal
surface, and $(\widehat{{\bf x}},\widehat{{\bf y}},\widehat{{\bf z}})$
represents the principal-axis system of the crystal ([100], [010], [001]).
For the (100) face we simply choose the $\widehat{{\bf x}}$ axis to lie
normal to the surface, the transformation matrix can be expressed by 
\begin{equation}
R_{1}=\left( 
\begin{array}{ccc}
0 & 1/\sqrt{2} & -1/\sqrt{2} \\ 
0 & 1/\sqrt{2} & 1/\sqrt{2} \\ 
1 & 0 & 0
\end{array}
\right) ,
\end{equation}
and for the (110) face crystal 
\[
R_{1}=\left( 
\begin{array}{ccc}
-1/\sqrt{2} & 1/\sqrt{2} & 0 \\ 
0 & 0 & 1 \\ 
1/\sqrt{2} & 1/\sqrt{2} & 0
\end{array}
\right) . 
\]
We can define a set of unit vectors for the incident light beam $\widehat{%
{\bf s}},$ $\widehat{{\bf k}},$and $\widehat{{\bf z}}$ such that $\widehat{%
{\bf s}}$ and $\widehat{{\bf k}}$ lie on the crystal face, perpendicular and
parallel to the plane of incidence, respectively, and $\widehat{{\bf z}}=%
\widehat{{\bf z}}^{\prime }$ is the surface normal, 
\begin{equation}
\left( 
\begin{array}{c}
\widehat{{\bf s}} \\ 
\widehat{{\bf k}} \\ 
\widehat{{\bf z}}
\end{array}
\right) =R_{0}\left( 
\begin{array}{c}
\widehat{{\bf x}} \\ 
\widehat{{\bf y}} \\ 
\widehat{{\bf z}}
\end{array}
\right) =\left( 
\begin{array}{ccc}
\sin \phi & -\cos \phi & 0 \\ 
\cos \phi & \sin \phi & 0 \\ 
0 & 0 & 1
\end{array}
\right) \left( 
\begin{array}{c}
\widehat{{\bf x}} \\ 
\widehat{{\bf y}} \\ 
\widehat{{\bf z}}
\end{array}
\right) .
\end{equation}
Here $\phi $ is the angle between $\widehat{{\bf k}}$ and $\widehat{{\bf x}}$%
, the azimuthal angle.

In many of the experiments, the single surface is rotated about its
azimuthal angle $\phi $ and the SH response is analyzed with respect to the
beam coordinates. Therefore, one needs to transform ${\bf {\chi }}^{(2)}$
from crystal coordinates into beam coordinates, as shown in figure 1, with
the appropriate transformation operations. The transformation rule for third
rank tensors is the following: 
\begin{equation}
\chi _{ijk}=%
\mathrel{\mathop{\sum }\limits_{l,m,n}}%
R_{il}R_{jm}R_{kn}\chi _{lmn},  \label{Eq-e}
\end{equation}
where $R=R_{0}R_{1}$ is the transformation operator from the crystal $(%
\widehat{{\bf x}},\widehat{{\bf y}},\widehat{{\bf z}})$ to beam $(\widehat{%
{\bf s}},\widehat{{\bf k}},\widehat{{\bf z}})$ coordinates.

Wide-gap zinc-blende II-VI\ semiconductors with symmetry $T_{d}(\overline{4}%
3m)$ are optically isotropic, but do not possess a center of inversion. The
bulk second-order susceptibility tensor has only one component, $\chi
_{xyz}=\chi _{yzx}=\chi _{zxy}=d$. Using the rule in Eq. (\ref{Eq-e}), this
tensor for the (100) face in beam coordinates can be written as 
\begin{equation}
{\bf {\chi }}_{bulk}^{(2)}(\phi )=\left( 
\begin{array}{cccccc}
0 & 0 & 0 & -\sin (2\phi )d & \cos (2\phi )d & 0 \\ 
0 & 0 & 0 & -\cos (2\phi )d & -\sin (2\phi )d & 0 \\ 
\cos (2\phi )d & -\cos (2\phi )d & 0 & 0 & 0 & -\sin (2\phi )d
\end{array}
\right) ,  \label{Eq-f}
\end{equation}
and for (111) face we have 
\begin{equation}
{\bf {\chi }}_{bulk}^{(2)}(\phi )=\left( 
\begin{array}{cccccc}
-\sqrt{\frac{2}{3}}\sin (3\phi )d & \sqrt{\frac{2}{3}}\sin (3\phi )d & 0 & 0
& -d/\sqrt{3} & -\sqrt{\frac{2}{3}}\cos (3\phi )d \\ 
-\sqrt{\frac{2}{3}}\cos (3\phi )d & \sqrt{\frac{2}{3}}\cos (3\phi )d & 0 & 
-d/\sqrt{3} & 0 & \sqrt{\frac{2}{3}}\sin (3\phi )d \\ 
-d/\sqrt{3} & -d/\sqrt{3} & 2d/\sqrt{3} & 0 & 0 & 0
\end{array}
\right) .
\end{equation}

The components of the fundamental field in the medium expressed in the beam
coordinate axes are 
\begin{equation}
\begin{tabular}{l}
$E_{s}t_{s},$ \\ 
$E_{k}=f_{c}E_{p}t_{p},$ \\ 
$E_{z}=f_{s}E_{p}t_{p},$%
\end{tabular}
{}
\end{equation}
where $f_{c,s}$ are the Fresnel factors and $t_{s,p}$ are the linear
transmission coefficients for the fundamental field, given by 
\begin{equation}
\vspace{0in}f_{s}=\frac{\sin \theta }{n(2\omega )},\quad
f_{c}=(1-f_{s}^{2})^{1/2},\quad t_{s}=\frac{2\cos \theta }{\cos \theta
+nf_{c}},\quad t_{p}=\frac{2\cos \theta }{n\cos \theta +f_{c}},  \label{Eq-g}
\end{equation}
with $f_{c}$ taken such that Im $f_{c}\geq 0,$ and Re $f_{c}\geq 0$ if Im $%
f_{c}=0$; $n(\omega )$ is the complex refractive index of the medium in
which the beam is propagating. $\theta $ is the angle of the beam
propagation in this medium. For \ the SH field, the coressponding $%
F_{s,}\,F_{c},\,T_{s},\,T_{p}$ are similar with refractive index $N(2\omega
) $ of SH light.

Through combining Eq. (\ref{Eq-d}) and Eqs. (\ref{Eq-f})-(\ref{Eq-g}), one
obtains the second-order polarization from the (100) face crystal, 
\begin{eqnarray}
P_{s}^{bulk} &=&-2d[\sin (2\phi )f_{c}f_{s}t_{p}^{2}E_{p}^{2}-\cos (2\phi
)f_{s}t_{p}t_{s}E_{p}E_{s}],  \nonumber \\
P_{k}^{bulk} &=&-2d[\cos (2\phi )f_{c}f_{s}t_{p}^{2}E_{p}^{2}+\sin (2\phi
)f_{s}t_{p}t_{s}E_{p}E_{s}], \\
P_{z}^{bulk} &=&d[\cos (2\phi )t_{s}^{2}E_{s}^{2}-\cos (2\phi
)f_{c}^{2}t_{p}^{2}E_{p}^{2}-2\sin (2\phi )f_{c}t_{p}t_{s}E_{p}E_{s}], 
\nonumber
\end{eqnarray}
and for the (111) crystal, 
\begin{eqnarray}
P_{s}^{bulk} &=&-\sqrt{\frac{2}{3}}d[\sin (3\phi )t_{s}^{2}E_{s}^{2}-\sin
(3\phi )f_{c}^{2}t_{p}^{2}E_{p}^{2}+2(\cos (3\phi )f_{c}+\sqrt{2}%
f_{s})t_{p}t_{s}E_{p}E_{s}],  \nonumber \\
P_{k}^{bulk} &=&-\sqrt{\frac{2}{3}}d[t_{s}^{2}\cos (3\phi )E_{s}^{2}+(\sqrt{2%
}f_{c}f_{s}-\cos (3\phi )f_{c}^{2})t_{p}^{2}E_{p}^{2}-2\sin (3\phi
)f_{c}t_{p}t_{s}E_{p}E_{s}], \\
P_{z}^{bulk} &=&-\frac{d}{\sqrt{3}}%
[t_{s}^{2}E_{s}^{2}-(f_{c}^{2}-2f_{s}^{2})E_{p}^{2}].  \nonumber
\end{eqnarray}

The SH fields generated by polarization are decomposed into {\it s- }and 
{\it p-}polarized components, 
\[
E_{s}^{bulk}=A_{s}\Omega L_{eff}P_{s}^{bulk} 
\]
and 
\begin{equation}
E_{p}^{bulk}=A_{p}\Omega L_{eff}[F_{s}P_{z}^{bulk}-F_{c}P_{k}^{bulk}]
\end{equation}
where $\Omega =2\omega /c$ is the magnitude of the wave vetor of the SH
light. $A_{s}$ and $A_{p}$ (given in Ref. \cite{Whubner94}) are independent
of the angle of rotation, but are dependent on the incident angle and the
optical frequency through a change in the index of refraction. The SH\
intensity is propotional to the absolute square of $E(2\omega )$, which for
either {\it s- }or {\it p-}polarized pump beam, from (100), (110) and (111)
face crystals are found from Table I (only the bulk susceptibility is
included).

\begin{center}
$
\begin{tabular}{c}
TABLE I. The second-harmonic field for bulk contibution \\ \hline\hline
\multicolumn{1}{l}{\bf (100) face} \\ 
\multicolumn{1}{l}{$E_{p,s}=-2d\,\Omega
\,L_{eff}\,f_{c}\,f_{s}\,t_{p}^{2}\cos (2\phi )A_{s}E_{p}^{2}$} \\ 
\multicolumn{1}{l}{$E_{p,p}=d\,\Omega
\,L_{eff}\,f_{c}\,t_{p}^{2}(2f_{s}\,F_{c}-f_{c}F_{s})\cos (2\phi
)A_{p}E_{p}^{2}$} \\ 
\multicolumn{1}{l}{$E_{s,s}=0$} \\ 
\multicolumn{1}{l}{$E_{s,p}=d\,\Omega \,F_{s}\,L_{eff}\,\,t_{s}^{2}\cos
(2\phi )A_{p}E_{s}^{2}$} \\ 
\multicolumn{1}{l}{\bf (110) face} \\ 
\multicolumn{1}{l}{$E_{p,s}=\frac{1}{2}d\,\Omega \,L_{eff}\,[(3\cos (2\phi
)-1)f_{c}^{2}-2f_{c}^{2}]\cos (\phi )A_{s}E_{p}^{2}$} \\ 
\multicolumn{1}{l}{$E_{p,p}=d\,\Omega \,L_{eff}\,[3\cos ^{2}\phi
\,f_{c}^{2}\,F_{c}-f_{s}^{2}F_{c}+2f_{c}\,f_{s}F_{s}]\sin (\phi
)A_{p}E_{p}^{2}$} \\ 
\multicolumn{1}{l}{$E_{s,s}=3\,d\,\Omega \,\cos \phi \sin ^{2}\phi
\,A_{s}E_{s}^{2}$} \\ 
\multicolumn{1}{l}{$E_{s,p}=\frac{1}{4}d\,\Omega \,L_{eff}\,F_{c}[\sin \phi
-3\sin (3\phi )]A_{p}E_{s}^{2}$} \\ 
\multicolumn{1}{l}{\bf (111) face} \\ 
\multicolumn{1}{l}{$E_{p,s}=\sqrt{\frac{2}{3}}\,d\,\Omega
\,L_{eff}\,f_{c}^{2}\,t_{p}^{2}\sin (3\phi )A_{s}E_{p}^{2}$} \\ 
\multicolumn{1}{l}{$E_{p,p}=\sqrt{\frac{1}{3}}\,d\,\Omega \,L_{eff}[\,\sqrt{2%
}\,f_{c}^{2}F_{c}\,\cos (3\phi
)+f_{c}^{2}\,F_{s}-2f_{c}\,f_{s}\,F_{c}-2f_{s}F_{s}]t_{p}^{2}A_{p}E_{p}^{2}$}
\\ 
\multicolumn{1}{l}{$E_{s,s}=-\sqrt{\frac{2}{3}}\,d\,\Omega
\,L_{eff}\,\,t_{s}^{2}\sin (3\phi )A_{s}E_{s}^{2}$} \\ 
\multicolumn{1}{l}{$E_{s,p}=\sqrt{\frac{1}{3}}\,d\,\Omega \,L_{eff}[\,\sqrt{2%
}\,F_{c}\cos (3\phi )-F_{s}]t_{s}^{2}A_{p}E_{s}^{2}$} \\ \hline\hline
\end{tabular}
$
\end{center}

The SH intensity is proportional to the absolute squre of $E(2\omega )$.
Figure 2 shows the rotation-angle dependence of the SH intensity $%
P_{in}-S_{out}$ for singular ZnSe(100) and ZnSe(111) with only the bulk
suscetptibility. For ZnSe(100), the SH\ intensity distribution shows a
fourfold symmetry, but for ZnSe(111) a full sixfold symmetry.

\subsection{Surface or Inrerface Contribution}

The surface or interface nonlinear susceptibility ${\bf \chi }_{surf}^{(2)}$
is a third rank tensor with 27 elements, which reflect the symetry of the
interface. For SHG the last two indices may be permuted at will, thus, the
nonlinear susceptibility tensor may have a maximum of 18 independent
nonvanishing elements( e.g. $C_{1}$), and higher symmetries lead to a
reduction in the number of independent and nonvanishing tensor elements.
Table II summarizes the results for the form of the surface nonlinear
susceptibility tensor ${\bf \chi }_{surf}^{(2)}$ for various symmetry
classes.

\begin{center}
$
\begin{tabular}{lcc}
\multicolumn{3}{l}{TABLE II. Independent nonvanishing elements of ${\bf \chi 
}_{surf}^{(2)}\,$ for continuous} \\ 
\multicolumn{3}{l}{point group for a surface in the $\widehat{{\bf x}}-%
\widehat{{\bf y}}${\bf \ }plane. Where mirror planes exist,} \\ 
\multicolumn{3}{l}{one of them is assumed to lie perpendicular to the $%
\widehat{{\bf y}}$-axis.\cite{Guyot86,Butcher90}} \\ \hline\hline
Point Group & $\;\;\,{}$ & Nonvanishing independent tensor elements \\ \hline
$C_{1}-1$ & \thinspace & \multicolumn{1}{l}{\it %
xxx,xyy,xzz,xyz=xzy,xzx=xxz,xxy=xyx,yxx,yyy,yzz,yyz=yzy,} \\ 
\thinspace & \thinspace \thinspace & \multicolumn{1}{l}{\it %
yzx=yxz,yxy=yyx,zxx,zyy,zzz,zyz=zzy,zzx=zxz,zxy=zyx} \\ 
$C_{s}-m$ & \thinspace & \multicolumn{1}{l}{\it %
xxx,xyy,xzz,xzx=xxz,yyz=yzy,yxy=yyx,zxx,zyy,zzz,zxz=zzx} \\ 
$C_{2}-2$ & $\,$ & \multicolumn{1}{l}{\it %
xyz=xzy,xxz=xzx,yxz=yzx,yzy=yyz,zxx,zyy,zzz,zxy=zyx} \\ 
$C_{2v}-mm2$ & $\,$ & \multicolumn{1}{l}{\it xxz=xzx,yyz=yzy,zxx,zyy,zzz} \\ 
$C_{3}-3$ & $\;\;\,$ & \multicolumn{1}{l}{\it %
xxx=-xyy=-yxy=-yyx,yyy=-yxx=-xxy=-xyx,} \\ 
\multicolumn{1}{c}{\thinspace} & $\,$ & \multicolumn{1}{l}{\it %
xxz=xzx=yyz=yzy,zxx=zyy,xyz=xzy=-yxz=-yzx,zzz} \\ 
$C_{3v}-3m$ & \  & \multicolumn{1}{l}{\it %
xxx=-xyy=-yxy=-yyx,xzx=xxz=yzy=yyz,zxx=zyy,zzz} \\ 
$C_{4}-4,C_{6}-6$ & \  & \multicolumn{1}{l}{\it %
xxz=xzx=yyz=yzy,zxx=zyy,xyz=xzy=-yxz=-yzx,zzz} \\ 
$C_{4v}-4mm$ & \thinspace & \multicolumn{1}{l}{\it %
xxz=zxx=yzy=yyz,zxx=zyy,zzz} \\ 
$C_{6v}-6mm$ & \multicolumn{1}{l}{\thinspace} & \multicolumn{1}{l}{\it %
xxz=zxx=yzy=yyz,zxx=zyy,zzz} \\ \hline\hline
\end{tabular}
$
\end{center}

We use the symble $\chi _{xyz}^{surf}$ to denote the surface(dipolar)
second-order susceptibility, and $P_{x,y,z}^{surf}$ for surface
polarization, e.g., 
\[
\begin{tabular}{llcccc}
\multicolumn{6}{l}{TABLE III. Rotation-angle dependences of the second-order
susceptibility tensor} \\ 
\multicolumn{6}{l}{components for different symmetry.} \\ \hline\hline
Surface & Tensor elements${}$ & $P_{in}-P_{out}$ & $P_{in}-S_{out}$ & $%
S_{in}-P_{out}$ & $S_{in}-S_{out}$ \\ 
\multicolumn{1}{c}{symmetry} & \multicolumn{1}{c}{(100) face {\it %
xyz\thinspace \thinspace }$^{a}$} & cos$(2\phi )$ & cos$(2\phi )$ & cos$%
(2\phi )$ & --- \\ 
\multicolumn{1}{c}{\thinspace \thinspace} & \multicolumn{1}{c}{(111) face 
{\it xyz\ }$^{a}$} & cos$(3\phi )$ & sin$(3\phi )$ & cos$(3\phi )$ & sin$%
(3\phi )$ \\ \hline
$C_{4v},C_{6v}$ & {\it xxz=yyz,zzz} & isotropic & --- & --- & --- \\ 
\thinspace & {\it zxx=zyy} & isotropic & --- & isotropic & --- \\ 
$C_{3v}\,^{b}$ & {\it xxx} & cos$(3\phi )$ & sin$(3\phi )$ & cos$(3\phi )$ & 
sin$(3\phi )$ \\ 
$C_{3}\,^{b}$ & $\,${\it xxx} & cos$(3\phi )$ & sin$(3\phi )$ & sin$(3\phi )$
& sin$(3\phi )$ \\ 
\thinspace & {\it yyy} & sin$(3\phi ),$sin$\phi $ & cos$\phi $ & cos$(3\phi
),$sin$\phi $ & cos$(3\phi ),$cos$\phi $ \\ 
$C_{2v}$ & $\,${\it zxx+zyy} & isotropic & --- & isotropic & --- \\ 
\thinspace & $\,${\it xxz+yyz} & isotropic & --- & --- & --- \\ 
\thinspace & {\it xxz-yyz} & sin$(2\phi )$ & sin$(2\phi )$ & --- & --- \\ 
\thinspace & z{\it xx-zyy} & sin$(2\phi )$ & --- & sin$(2\phi )$ & --- \\ 
\thinspace & z{\it zz} & isotropic & --- & --- & --- \\ 
$C_{s}\,^{b}$ & {\it xxx,xyy,yyx} & cos$(3\phi ),$cos$\phi $ & sin$(3\phi ),$%
sin$\phi $ & cos$(3\phi ),$cos$\phi $ & sin$(3\phi ),$sin$\phi $ \\ 
\multicolumn{1}{c}{\thinspace} & {\it xxz,yyz} & cos$(2\phi )$ & sin$(2\phi
) $ & --- & --- \\ 
\thinspace & {\it xzz} & cos$\phi $ & sin$\phi $ & --- & --- \\ 
$C_{1}\,^{b}$ & {\it xxx,xyy,yyx} & cos$(3\phi ),$cos$\phi $ & sin$(3\phi ),$%
sin$\phi $ & cos$(3\phi ),$cos$\phi $ & sin$(3\phi ),$sin$\phi $ \\ 
\thinspace & {\it yyy,xyy,yyx} & sin$(3\phi ),$sin$\phi $ & cos$(3\phi ),$cos%
$\phi $ & sin$(3\phi ),$sin$\phi $ & cos$(3\phi ),$cos$\phi $ \\ 
\thinspace & {\it xyz,yxz} & sin$(2\phi )$ & cos$(2\phi )$ & --- & --- \\ 
\thinspace & {\it zxy} & sin$(2\phi )$ & --- & sin$(2\phi )$ & --- \\ 
\thinspace & {\it xzz} & sin$\phi $ & sin$\phi $ & --- & --- \\ \hline\hline
\multicolumn{6}{l}{$^{a}${\small \ Only the bulk susceptibility is included.}
} \\ 
\multicolumn{6}{l}{$^{b}${\small \ The isotropic terms are not shown.}}
\end{tabular}
\]
\begin{equation}
P_{x}^{surf}=\chi _{xyz}^{surf}E_{y}E_{z},\,etc.,
\end{equation}
in the $(x,y,z)$ system. The transformation rule Eq. (\ref{Eq-e}) is used to
transfer the tensor $\chi _{xyz}^{surf}$ from crystal coordinates to beam
coordinates $\chi _{skz}^{surf}$. Following Sipe \cite{Sipe87}, the SH field
induced by the sheet of polarization can be written as 
\begin{equation}
\begin{tabular}{l}
$E_{s}^{surf}=A_{s}\Omega P_{s}^{surf},$ \\ 
$E_{p}^{surf}=A_{p}\Omega \lbrack F_{s}\varepsilon (2\omega
)P_{z}^{surf}-F_{c}P_{k}^{surf}],$%
\end{tabular}
\end{equation}
where $\varepsilon (2\omega )$ is the dielectric constant at frequency $%
2\omega $.

We assume for simplicity that the surface has a simple unreconstructed
structure, and thus for a particular face it has the same symmetry as the
bulk. Therefore we use $C_{3v}$ symmetry for the (111) surface and $C_{4v}$
symmetry for the (100) face \cite{Sipe87}. In fact, there exist many local
microscopic structures, even in a nominally single surfaces, and in such
cases, the microscopic symmetry of the structures, such as monoatomic steps,
etc. would allow some specific tensor element to exist. That is, we may
observe a lower symmetry than expected if the newly introduced tensor is of
different symmetry. Table III summarizes the results in cases where surface
susceptibility is presented.

From Table III, we see that there are always plural tensor elements which
cause the same rotational symmetry. Therefor, the determination of the
relevant tensor elements must depend on other sources of information, which
may be obtained experimentally. Note that because SHG is described by a
third rank tensor, all surface symmetries higher than $C_{3v}$ yield an
isotropic response.

\section{EXPERIMENTAL METHODS}

The experimental setup for SHG measurement is shown in Fig. 3. The light
source of the fundamental frequency is a picosecond mode-locked Nd:YAG
(yttrium aluminum garnet) laser (Light Conversion Ltd. EKSPLA 2143A) with
light pulses of \ wavelength 1.06$\mu $m, duration time 20 ps, and repetion
rate 10 Hz. The linear polarization of the fundamental beam is rotated to
the desired angle by the combination of the half wave plate ($\lambda /2$)
and the input Glan-Taylor polarizer prism P1. The polarized beam is focused
into a spot of 3 mm diameter on the sample surface at an incident angle of $%
45%
{{}^\circ}%
$. In order to avoid damaging the sample, the pulse energy was set to about
0.5-0.3mJ per pulse throughout the experiment. The visible-cut filter F1
situated immediately before the sample surface removes the visible light
component, especailly SH signals which are generated in the preceding
polarizer and lens. The reflected fundamental beam and the SH signal from
the surfaceare are collinear beyond the sample surface, and the former is
eliminated from the beam by an IR-cut filter F2 and an interference filter
F3 (Melles Griot 03FIB008, $\lambda =550$ nm, FWHM=70 nm). The linear
polarization of the SH signal to be measured by the detection system was
selected by rotation of the analyzer P2. The signal was finally detected by
a photomultiplier (Hamamatsu R1464). About 4\% of the reflected laser beam
is directed through the reference arm containing a crystalline quartz plate.
The SH\ intensity from quartz plate can be used for providing a reference to
remove laser intensity fluctuations.This signal is detected by another
photomultiplier (Hamamatsu R928) through an interference filter F3. Both
electronics signals from the two PMT outputs were fed into a BOXCAR (SRS
250) averaged over 50 pulses and then recorded using a computer.

The ZnSe(100)\ and ZnSe(111) crystals grown by the Markov method were
commericailly purchased from Ma Teck Material Technologie \&\ Kristalle
GmbH, Germany \cite{Mateck}. Their surfaces are chemical and mechanically
polished. The samples were mounted on a rotation stage, with the surface
normal set parallel to the rotating axis of the stage and could be rotated
freely.

\section{RESULTS AND DISCUSSION}

In Figure 4(a)-(d), the intensity from ZnSe(100) is shown as a function of
the azimuthal orientation for the four different polarization combinations: $%
P_{in}-P_{out},\,P_{in}-S_{out},\,S_{in}-P_{out},S_{in}-S_{out}$. All
experiments were made under the same conditions which makes it possible to
compare the intensities of the different combinations. Fig. 4(a) [Fig. 4(b)]
displays the the $p$- [$s$-]polarized SH intensity $I_{pp}$ [$I_{sp}$] with $%
p$-[$s$-] polarizedincident light, whereas Fig. 4(c) [Fig. 4(d)] shows the $%
s $-polarized SH intensity $I_{ps}$ [$I_{ss}$] with $p$-[$s$-] polarized
incident light. All plots clearly exhibit twofold symmetry. It seems to be
consistent with a $C_{s}$ symmetry. A more detailed analysis of the
experimental reuslts (see below) shows the interference of the surface $%
C_{s} $ symmetry with the underlying $C_{4v}$ symmetry of the ZnSe(100). The
data for the four cases are fit by the trigonometric function square plus a
constant, this small constant can be due to light at other frequencies
leaking through the filters. We might be able to obtain relative values for
the surface and bulk contributions if the function of the incident angle
multiplying the different surface and bulk contributions would change
significantly with $\theta $ (see Eq. (\ref{Eq-g}) and TABLE I ). However,
with the experimental conditions used here, changing from normal to grazing
incidence decreases $f_{c}$ and $F_{c}$ by $5\%,$ whereas $f_{s}$ and $F_{s}$
remain small. Hence any attempt to seperate bulk and surface contribution by
the use of this method seams to be inappropriate. It is for this reason that
in all experiments a fixed angle of incidence of $45%
{{}^\circ}%
$ was used.

For simplicity we concentrate on the $s$-polarized SHG response for an $s$%
-polarized pump beam, becuase this signal only contains anisotropic terms
and will be most sensitive for the surface symmetry. This conclusion arises
because the bulk contribution is forbidden, as seen in\ Table III. According
to \cite{Sipe87}, for the (100) face, we see that the surface in general has 
$C_{4v}$ symmetry. However, such a mechanism should lead no $\phi $-angle
dependence. Even though we replace the symmetry$\ C_{4v}$ with lower
symmetry $C_{2v}$, we cannot obtain any $\phi $-angle dependent signal. Thus
we must allow the effective symmetry of the surface to lower from $C_{2v}$
to $C_{s}$, corresponding to the observed symmetry of the rotation angle
dependence. From Table III, for $C_{s}$ symmetry only the following
components of surface or interface nonlinear susceptibility do not vanish 
\begin{equation}
\chi _{xxx},\chi _{xyy},\text{and }\chi _{yxy}=\chi _{yyx}.
\end{equation}
And a similar experssion for the $S_{in}-S_{out}$ case is derived as 
\begin{equation}
E_{s,p}(2\omega )/E_{s}^{2}(\omega )A_{p}\varpropto \chi _{xxx}\,\sin
^{3}\phi -(\chi _{xyy}+2\,\chi _{yyx}\,)\,\sin \phi \,\cos ^{2}\phi \,\,.
\end{equation}
The solid line in Fig. 4(d) is a least-squares fit of SHG intensity to the
data, from which we obtain the ratio 
\begin{equation}
\frac{\chi _{xxx}}{\chi _{xyy}+2\,\chi _{yyx}\,}=0.276\,,
\end{equation}
the agreement with experimental reuslt justifies our choice of $C_{s}$
symmetry. There is a large scatter in the data for $S_{in}-S_{out}$
polarization combination, because its absolute SH intensity is small, thus
we cannot obtain a complete fit to every data point in this configuration.
In Ref. \cite{Tepper89}, value of $\chi _{xxx}\diagup (\chi _{xyy}+2\,\chi
_{yyx})=1.10$ was obtained from the NaCl(100)-air interface. This difference
is not surprising given the fact that the absorption depth in the oxide
layers of the NaCl(100) is more that of ZnSe(100) in our experiment.

This $C_{s}$ symmetry may be caused by (1) the interface between oxide
absorbate and sample \cite{Armstrong92,Tepper89}, (2) a nonideal cut of the
crystal, the domain formation with different symmetry axis \cite{Bottomley93}%
. When one only consider the bulk susceptibility, $\phi $ dependence for $%
S_{in}-S_{out}$ is very sensitive to the miscut angle of the crystal axis,
even its absolute SH intensity is very small. If we consider that the
surface tensor elements $\chi _{xxx},\chi _{xyy},$and $\chi _{yxy}$ have
relevance to the miscut angle of ZnSe(100) crystal, which is less than $0.5%
{{}^\circ}%
$ \cite{Mateck}, the response of $S_{in}-S_{out}$ is determined to be more
than an order of magnitude smaller than other corresponding cases \cite
{Takebayashi97,Yamada94}. C. Yamada and K. Kimura \cite{Yamada93} found it
was very difficult to measure $S_{in}-S_{out}$ rotational-angle dependence
because the signal was poor under UHV condition. However, we observed that
the output was about half of the other cases. So we concluded that in our
experiment the interface plays an impoartant role for producing the
anisotropic in $S_{in}-S_{out}$ configuration. This is supported in part by
similar measurements made for GaAs(100) surfaces contaminated with carbon
and oxygen which revealed a high degree of anisotropy in the SHG signal \cite
{Pemble92}. The anisotropiy cannot be ascribed to any particular surface
symmetry. One possible explanation may be suggested: the anisotropy may
arise due to variations in oxide layer thickness which could be regarded as
generating a stepped buried ZnSe(100)/oxide layer interface.

For three configurations ($P_{in}-P_{out}$, $S_{in}-P_{out}$, $%
P_{in}-S_{out} $), the SHG intensity distribution shows a $C_{4v}$ symmetry
if we only consider the bulk susceptibility. Further, we notice that the
oxide layer interface would lower the effective symmetry of the surface from 
$C_{4v}$ to $C_{2v}$. In Fig 4.(a)-(c), we observe twofold symmetries not
fourfold for three configurations, which is the result of interference of
bulk SHG with surface SHG. Threfore, the observed rotation-angle dependence
can be described as 
\begin{equation}
I_{s.p}\propto \left| A+B\cos (2\phi )+C\sin (2\phi )\right| ^{2},
\label{Eq-h}
\end{equation}
where $A$ represents the isotropic surface contribution, $B$ is the bulk
dipole term and $C$ the surface anisotropic term. We least-square fitted the
observed rotation-angle dependence for three configurations by Eq. (\ref
{Eq-h}) with $A,$ $B$ and $C$ used as adjustable parameters. These fitting
parameters are summarized in Table IV for the SHG data. As shown in Fig.
4(a)-(c), the fit was satisfactory.

\begin{center}
TABLE IV. Values obtained for fitting parameters in Eq. (\ref{Eq-h})

\begin{tabular}{ccccccc}
\hline\hline
Configuration &  & A &  & B &  & C \\ \hline
$P_{in}-P_{out}$ &  & \thinspace -0.034 &  & \thinspace\ 0.203 &  & 1.201 \\ 
$P_{in}-S_{out}$ &  & \thinspace 0.046 &  & 1.200 &  & -0.126 \\ 
$S_{in}-P_{out}$ &  & \thinspace 0.062 &  & 0.039 &  & 1.150 \\ \hline\hline
\end{tabular}
\end{center}

In the $P_{in}-P_{out}$ combination (Fig 5.), the peak at $\phi =135%
{{}^\circ}%
$ is higher than that at $\phi =45%
{{}^\circ}%
,$ and the peak at $\phi =315%
{{}^\circ}%
$ is higher than that at $\phi =225%
{{}^\circ}%
.$ These diffrences arise from the interference between the isotropic
surface SHG and the bulk SHG. The signs of the nonlinear susceptibility
components for the exciting electric fields in the (011) and (0$\overline{1}$%
1) planes are opposite. This is because the bulk Zn-Se-Zn-Se- chain along
the [011] has Zn atoms higher than Se atoms and Zn-Se-Zn-Se- chain alone [0$%
\overline{1}$1] has Se atoms higher than Zn atoms. This phase difference of $%
180%
{{}^\circ}%
$ between the electronic wave functions of the two bulk chains leads to the
diffrence in the interference between the bulk and surface SHG and to the
change of the SHG peak intensities.

\section{SUMMARY}

We have developed a phenonmenological model and performed experiments to
determine the symmetry of the noncentrosymmetric semiconductor crystal
ZnSe(100) covered with an oxide layer. Under the four combinations of
fundamental and harmonic linear polarization states considered, the harmnic
intensity can be experessed as a function of the light polarization and as a
function of the azimuthal angle. Using the caclulated sensitivity to
rotation angle of the SHG\ signal from the bulk and surfaces, we have been
able to deduce the symmetry of the noncentrosymmetric crystal surfaces. We
found that the measurement using $S_{in}-S_{out}$ is particularly useful in
determining the symmetry of the oxdized layer interface, which would lower
the effective symmetry of the surface from $C_{4v}$ to $C_{2v}.$ The
separation between bulk and interface or surface SHG\ demonstrated here is
promising for the application of this technique to the study of surface and
interfacial properties. This way can be used to detect the quality of the
substrate surface for growth of II-VI epilayers on ZnSe layers.

In addition, we have shown that the [011] and [0$\overline{1}$1] directions
can be distinguished through the analysis of $p$-incident and $p$-output
confugration.

\begin{figure}[tbp]
\caption{Principal geometry for SHG in reflection. In the experiment the
surface is rotated its normal Z (azimuthal angle $\protect\phi $ ), and the
polarization of both fundamental and SH can be varied to any direction
between $s$ and $p.$}
\label{F1}
\end{figure}

\begin{figure}[tbp]
\caption{The rotation-angle dependence of the SH intensity $P_{in}-S_{out}$
for singular ZnSe(100) and ZnSe(111) with only the bulk suscetptibility.}
\label{F2}
\end{figure}

\begin{figure}[tbp]
\caption{Schematic diagram of the experimental configuration used to measure
the SHG from ZnSe single crystal surfaces.}
\end{figure}

\begin{figure}[tbp]
\end{figure}

\begin{figure}[tbp]
\caption{Rotational anisotropy of SHG from ZnSe(100) surface in free air for
a set of input and output polarization combinations. The circles are the
experimental points, and the solid lines are least-squares fit to
theoretical calculation.}
\end{figure}

\begin{figure}[tbp]
\caption{Polar plot of the SHG\ intensity from ZnSe(100). The pump and SHG
beam were both $p$ polarized.}
\end{figure}

\begin{acknowledgement}
We would like to thank Dr. Yudan Cheng for helpful discussion, Hao Yin from
Physics Department at Uppsala Univ. for technical assistance, and Dr. Emad
Mukhtar from Department of Physical Chemistry at Uppsala Univ. for BOXCAR\
equipment. Wolfgang Seibt, Dr. Duc Tran Chinh and Bo Carman are gratefully
acknowledged.
\end{acknowledgement}

\end{document}